\documentclass[conference]{IEEEtran}
\IEEEoverridecommandlockouts

\usepackage{cite}
\usepackage{amsmath,amssymb,amsfonts}
\usepackage{algorithmic}
\usepackage{graphicx}
\usepackage{textcomp}
\usepackage{xcolor}
\usepackage{soul}
\usepackage{booktabs}
\usepackage{tabularx}
\usepackage{array}
\usepackage{xurl}
\usepackage{listings}
\makeatletter
\renewcommand{\lst@makecaption}[2]{%
    \def\@captype{lstlisting}%
    \@makecaption{#1}{#2}%
    \vspace{4pt}%
}
\makeatother

\usepackage[hidelinks]{hyperref}

\def\BibTeX{{\rm B\kern-.05em{\sc i\kern-.025em b}\kern-.08em
    T\kern-.1667em\lower.7ex\hbox{E}\kern-.125emX}}

\begin{document}

\title{Embedding Drift in Code Vulnerability
Models Under Intended Behaviour-Preserving Transformations\\}

\author{
\IEEEauthorblockN{
Hasti Ghaneshirazi, Tahsin Reza, and Ladan Tahvildari
}
\IEEEauthorblockA{
\textit{Electrical and Computer Engineering} \\
\textit{University of Waterloo} \\
Waterloo, Canada \\
\{hghanesh,tahsin.reza,ladan.tahvildari\}@uwaterloo.ca
}
}


\maketitle

\begin{abstract}
Code edits designed to preserve intended behaviour can shift frozen code
embeddings across classifier decision boundaries, causing correctly detected vulnerabilities to be predicted as benign. This instability is important because harmless changes such as removing comments, adding code that never runs, renaming variables, or rewriting a loop should not alter a model's security judgment. We study this problem using 15,000 C/C++ functions organized as 7,500 vulnerable-patched Big-Vul pairs. We apply four mutation tracks to both classes: removing comments, adding code that never runs, renaming variables and rewriting loops, and combining all changes. We then evaluate frozen \texttt{microsoft/codebert-base} embeddings with six classifiers and introduce a train-only defense that projects clean and mutated versions closer together while preserving vulnerability-class information. Under combined mutations, the baseline Vulnerable Flip Rate (VFR) ranges from 35.55\% to 42.15\%. The defence reduces the mean common-set VFR from 22.60\% to 11.32\% and increases mean mutated accuracy from 55.33\% to 57.10\%. For Logistic Regression, VFR decreases by 17.29 percentage points, although benign-to-vulnerable flips increase by 11.95 percentage points. The defence therefore improves robustness but introduces a false-positive trade-off and does not fully eliminate prediction instability.
\end{abstract}

\section{Introduction}

Learning-based code vulnerability models are increasingly developed to
support automated security review at the function and line
levels~\cite{zhou2019devign,fu2022linevul}. Their predictions, however,
should remain stable when a program is rewritten without changing its
intended behaviour. Prior work has shown that semantics-preserving
source transformations, identifier substitutions, and dead-code
insertion can cause code models to change otherwise correct
predictions~\cite{henkel2022semantic,yang2022alert,na2023dip}. For
example, removing a comment, renaming a variable, adding code that never
runs, or rewriting a loop should not cause a correctly detected
vulnerability to be classified as benign. Such failures are important
because developers may trust the changed prediction even though the
underlying security problem remains unchanged.

This instability can occur because pretrained code models encode
surface-level, syntactic, structural, and semantic information to
different degrees~\cite{karmakar2021pretrained}. Their representations
may therefore remain sensitive to changes in program form even when the
intended functionality is preserved
~\cite{jain2021contrastive,henkel2022semantic}. A harmless
transformation may shift a function's embedding across a downstream
classifier's decision boundary and alter its predicted security
label~\cite{yang2022alert,na2023dip}. Standard clean-set accuracy does
not reveal this problem because it does not compare predictions for
clean and transformed versions of the same function. We focus
particularly on the \textit{Vulnerable Flip Rate} (VFR), which measures
how often vulnerabilities correctly detected in clean code become
benign predictions after mutation.

To study this problem, we construct a paired benchmark of 15,000 C/C++
functions from Big-Vul, containing 7,500 vulnerable functions and their
patched counterparts~\cite{fan2020bigvul}. We apply four deterministic
mutation tracks to both classes: removing comments, inserting code that
never runs, renaming variables and rewriting loops, and combining all
transformations. Frozen \texttt{microsoft/codebert-base}
embeddings~\cite{feng2020codebert} are then evaluated using six
downstream classifier families under fixed, leakage-resistant training,
validation, and test splits.

Motivated by prior work on robust and transformation-invariant code
representations~\cite{henkel2022semantic,jain2021contrastive}, we
introduce a train-only projection defence that encourages clean and
mutated versions of the same function to remain close while retaining
information needed for vulnerability classification. This objective
requires an explicit trade-off: optimizing too strongly for invariance,
particularly through a large dimensional reduction 
bottleneck, may wash out subtle security-relevant details, such as a
missing bounds check, that distinguish vulnerable code from its safe
patch. The defence therefore combines an invariance objective with a
classification objective rather than minimizing representation
distance alone. It is evaluated using clean and mutated accuracy,
macro-F1, overall prediction flips, VFR, benign-to-vulnerable flips,
paired statistical tests, grouped bootstrap confidence intervals, and
PCA and UMAP visualizations~\cite{jolliffe2016pca,mcinnes2018umap}.

This study makes three main contributions. First, it provides a
controlled evaluation of four mutation types across six classifier
families. Second, it proposes a learned projection defence that reduces harmful vulnerable-to-benign prediction changes overall while explicitly
characterizing the accompanying false-positive trade-off; the
defence is trained only on the training split, selected using validation
data, and evaluated on a fully held-out test split. Third, it presents a
bidirectional robustness analysis that captures both missed
vulnerabilities and increased false-positive risk.

\begin{figure*}[t]
\centering
\includegraphics[width=\textwidth]{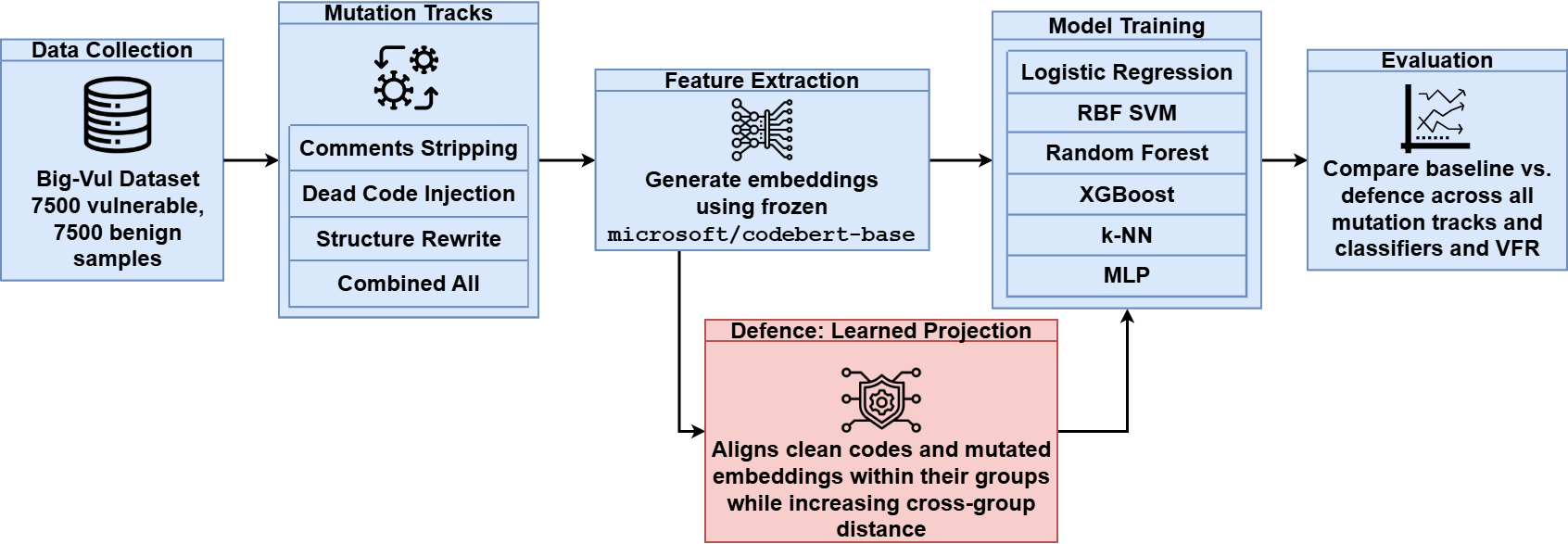}
\caption{Experimental workflow and system design. Frozen CodeBERT
generates 768-dimensional baseline embeddings from clean and mutated
code. The defence trains projection networks with latent dimensions of
128, 256, and 512 using the training split, selects the final
256-dimensional configuration using validation data, and applies it to
the held-out test split. Baseline and defended classifiers are compared
across mutation tracks using performance and robustness metrics,
including VFR.}
\label{fig:diagram}
\end{figure*}

\section{Background and Related Work}

\paragraph{Code representations and vulnerability detection.}

Learning-based vulnerability detectors encode source code into vector
representations for downstream classification. Early systems such as
Devign modelled syntax, control flow, and data dependencies as
graphs~\cite{zhou2019devign}, while later approaches adopted
transformers pretrained on large code corpora. CodeBERT learns joint
code and natural-language representations~\cite{feng2020codebert}, and
LineVul applies transformer features to function-level vulnerability
detection and vulnerable-line localization~\cite{fu2022linevul}.
However, encoding lexical, syntactic, structural, and semantic
information does not guarantee that predictions rely on the intended
program behaviour~\cite{karmakar2021pretrained}. This study therefore
uses frozen CodeBERT embeddings with Logistic Regression, SVM, Random
Forest, XGBoost, $k$-NN, and MLP classifiers to examine whether
non-functional changes move samples across downstream decision
boundaries.

\paragraph{Behaviour-preserving transformations and robustness.}

Code models should remain stable under transformations that preserve
intended program behaviour. Prior work has reported instability under
semantics-preserving transformations~\cite{henkel2022semantic},
identifier substitutions~\cite{yang2022alert}, and dead-code
insertion~\cite{na2023dip}, suggesting that predictions may depend on
lexical or structural cues that are not central to program semantics.
This study evaluates four deterministic mutation tracks: comment
removal, insertion of code that never runs, identifier renaming and loop
rewriting, and their combined application. Unlike adaptive attacks, the
same transformation procedure is applied to every eligible vulnerable
and patched sample, enabling controlled comparison of embedding
displacement and mutation-induced prediction changes.

\paragraph{Paired robustness evaluation.}

Clean and mutated accuracy or macro-F1 do not reveal which predictions
changed or the direction of those changes. Similar aggregate
performance may therefore conceal substantial sample-level
instability. Vulnerable-to-benign transitions increase the risk of
missed vulnerabilities, whereas benign-to-vulnerable transitions
increase false-positive risk. This study supplements conventional
performance measures with paired prediction disagreement and
class-conditional transition rates. The \emph{Vulnerable Flip Rate}
(VFR) measures the proportion of vulnerabilities correctly detected in
clean code that become benign predictions after mutation, while the
reverse transition measures additional false-positive risk. This
paired analysis complements recent work emphasizing more realistic
evaluation practices for vulnerability detection~\cite{ding2025primevul}.

\paragraph{Dataset quality and split integrity.}

The benchmark is derived from Big-Vul, which links vulnerable C/C++
functions to vulnerability-fixing commits~\cite{fan2020bigvul}.
Although this structure supports vulnerable--patched pairing, such
datasets can contain duplicate code, noisy labels, incomplete fixes,
and tangled commits in which one change addresses multiple files or
distinct issues~\cite{croft2023dataquality}. In these cases, the
\texttt{func\_after} version may not represent a fully benign function,
or the observed code difference may include changes unrelated to the
target vulnerability. These factors can weaken the correspondence
between a pair and its assigned labels.

The benchmark construction removes records with empty or identical
before-and-after functions and preserves each vulnerable function with
its patched counterpart. It also groups pairs, related samples, and
exact duplicates so that they cannot cross the training, validation,
and test splits. Mutations are applied symmetrically to both classes to
prevent mutation presence from becoming a label-correlated shortcut.
These controls reduce leakage and obvious inconsistencies, but they
cannot fully remove label noise originating from multi-vulnerability
commits, incomplete patches, or unrelated edits. Consequently, the observed clean accuracy of approximately
53--59\% across classifiers may reflect both limitations of the frozen
representation and an upper bound imposed by benchmark ambiguity. It should therefore not be interpreted solely as a
classifier or embedding-performance ceiling. PrimeVul similarly
highlights the importance of de-duplication, realistic splitting, and
careful benchmark construction~\cite{ding2025primevul}.

\paragraph{Robust representations and lightweight defences.}

Prior work improves robustness through contrastive
pretraining~\cite{jain2021contrastive}, transformation-based
augmentation~\cite{wang2022bridging}, adversarial
training~\cite{henkel2022semantic}, and causal learning that reduces
reliance on spurious features~\cite{rahman2024causalvul}. The defence used in this study is a smaller intervention: CodeBERT
remains frozen while a
$768\!\rightarrow\!512\!\rightarrow\!d$ projection network, where
$d\in\{128,256,512\}$, is trained with joint classification and
clean--mutation invariance objectives. The classification term encourages the projection to retain
vulnerability-related information, while the invariance term encourages
clean and mutated versions of the same function to remain close.

This design also introduces a trade-off. A projection that emphasizes
invariance too strongly may suppress subtle local evidence, such as a
missing bounds check, that distinguishes vulnerable code from its
patched counterpart. The projected representations are therefore
evaluated not only by reduced VFR, but also by clean and mutated
accuracy, macro-F1, overall prediction flips, and benign-to-vulnerable
transitions. The same six classifier families are retrained on the
defended representations to determine whether improved stability is
achieved without simply shifting predictions toward one class.

Overall, this work combines frozen code representations, deterministic
multi-track robustness testing, paired evaluation, split-integrity
controls, explicit consideration of benchmark noise, and lightweight
invariance learning within one controlled experimental pipeline.
\begin{figure*}[t]
\centering
\includegraphics[width=\textwidth]{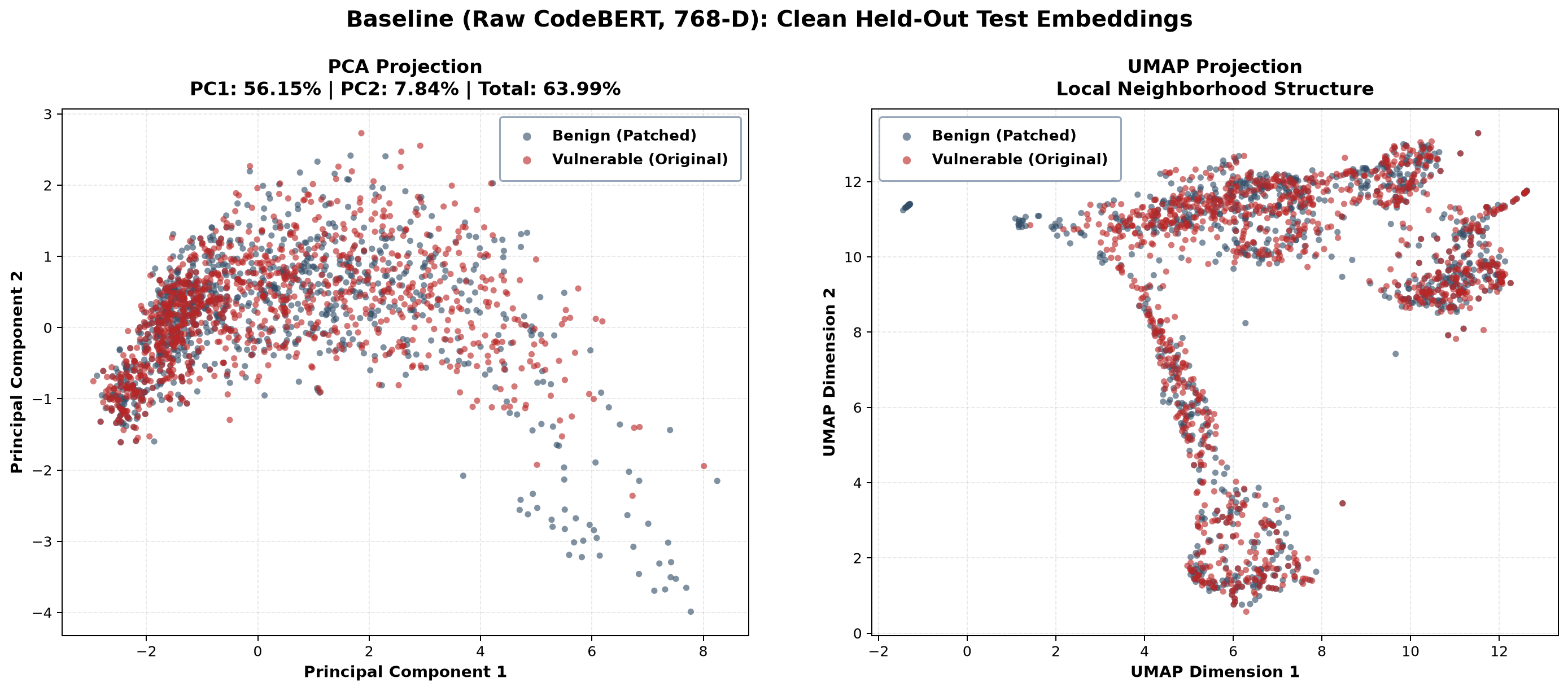}
\caption{PCA and UMAP views of held-out clean test embeddings in the frozen baseline CodeBERT space. The projections are qualitative and do not determine the classifier metrics.}
\label{fig:baseline_global_separation}
\end{figure*}

\section{Approach}

The methodology is guided by four research questions that frame the
study's evaluation and mitigation objectives.

\noindent\textbf{RQ1: How much prediction instability is hidden by
aggregate performance metrics?} This question examines whether clean and
mutated inputs receive different predictions even when their overall
accuracy and macro-F1 scores remain similar.

\noindent\textbf{RQ2: How sensitive are frozen code representations to
intended behaviour-preserving edits?} This question investigates whether
non-functional code transformations alter embedding geometry and
downstream vulnerability predictions.

\noindent\textbf{RQ3: How can robustness be evaluated while controlling
for dataset integrity and mutation-related shortcuts?} This question
examines how paired samples, related records, duplicate code, and
mutation artifacts should be controlled to support a reliable
evaluation.

\noindent\textbf{RQ4: Can a lightweight defence improve prediction
stability without introducing unacceptable false-positive risk?} This
question evaluates whether an invariance-oriented projection can reduce
vulnerable-to-benign flips while preserving clean performance and
limiting benign-to-vulnerable transitions.

Figure~\ref{fig:diagram} summarizes the experimental workflow and system
design. The pipeline begins with paired vulnerable and patched functions,
followed by deterministic mutation generation, dataset splitting, and
frozen CodeBERT embedding extraction. The resulting 768-dimensional
embeddings are used for baseline classifier evaluation. In the defended
setting, projection networks with latent dimensions of 128, 256, and 512
are trained and compared using only the training and validation splits.
The final 256-dimensional configuration is selected through a predefined
validation criterion and then applied unchanged to the held-out test
split across three projector seeds. Baseline and defended classifiers
are evaluated across all mutation tracks using paired robustness and
statistical analyses. The following subsections describe the benchmark
construction, mutation procedure, dataset splitting, representation
extraction, classifier evaluation, defence training, dimensional
ablation, defended embedding extraction, classifier retraining, and
statistical analysis in greater detail.

\smallskip
\noindent\textbf{Step 1: Paired Benchmark Construction}

The benchmark is constructed from the \texttt{train} split of the
Hugging Face dataset \texttt{bstee615/bigvul}. We retain vulnerability
records whose \texttt{vul} field equals 1 and for which both
\texttt{func\_before} and \texttt{func\_after} are present and non-empty.
Records whose before- and after-patch functions are identical after
removing surrounding whitespace are discarded, and exact duplicate
before--after function pairs are removed. From the remaining records,
7,500 patch pairs are sampled without replacement using random seed 42.
Each record contributes two linked samples: \texttt{func\_before} is
labelled vulnerable and \texttt{func\_after} is treated as its patched
benign counterpart. The resulting balanced benchmark contains 15,000
C/C++ functions. Each pair receives a stable \texttt{pair\_id}. A
\texttt{split\_group\_id} is assigned using the CVE identifier when
available, the project and commit identifier otherwise, and the pair
identifier as a fallback. Both members of each pair share the same
grouping identifiers, and every mutation track is applied to both
vulnerable and patched samples.

\begin{figure*}[t]
\centering
\includegraphics[width=\textwidth]{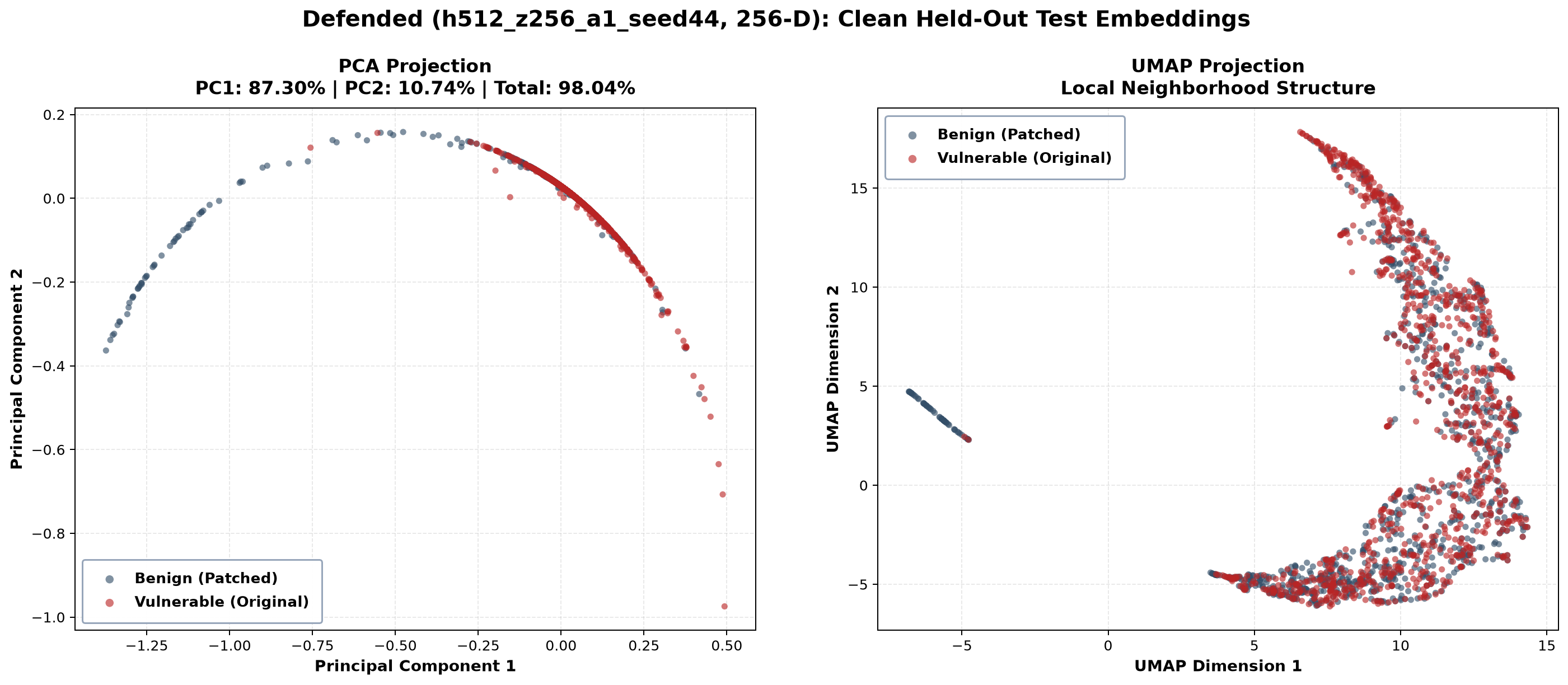}
\caption{PCA and UMAP views of held-out clean test embeddings in the selected 256-dimensional defended space. For each displayed run, the reducers are fit on defended training embeddings from the same projector seed and then applied to the corresponding test embeddings.}
\label{fig:defended_global_separation}
\end{figure*}

\begin{table*}[t]
\centering
\small
\caption{Test-Set Multi-Track Mutation Ablation Matrix across Baseline Embedding Classifiers}
\label{tab:ablation_matrix}
\begin{tabular}{llccccc}
\toprule
\textbf{Classifier Model} & \textbf{Mutation Track} &
\textbf{C-Acc (\%)} & \textbf{C-F1 (\%)} &
\textbf{M-Acc (\%)} & \textbf{M-F1 (\%)} &
\textbf{VFR (\%)} \\ \midrule

MLP Classifier & Combined All
& 55.56 & 55.30 & 53.87 & 52.74 & \textbf{37.41} \\
(128, 64) & Comments Only
& 55.56 & 55.30 & 54.98 & 54.93 & 3.89 \\
& Dead Code Only
& 55.56 & 55.30 & 54.18 & 52.99 & 28.33 \\
& Structure Only
& 55.56 & 55.30 & 55.60 & 55.28 & 20.37 \\ \midrule

Logistic Regression & Combined All
& 58.53 & 58.50 & 56.36 & 56.13 & \textbf{35.55} \\
& Comments Only
& 58.53 & 58.50 & 57.56 & 57.44 & 6.94 \\
& Dead Code Only
& 58.53 & 58.50 & 57.38 & 57.33 & 18.79 \\
& Structure Only
& 58.53 & 58.50 & 57.16 & 56.95 & 25.87 \\ \midrule

SVM (RBF Kernel) & Combined All
& 59.42 & 59.23 & 56.53 & 55.94 & \textbf{41.07} \\
& Comments Only
& 59.42 & 59.23 & 57.87 & 57.01 & 3.22 \\
& Dead Code Only
& 59.42 & 59.23 & 58.53 & 58.30 & 24.03 \\
& Structure Only
& 59.42 & 59.23 & 58.58 & 58.48 & 21.48 \\ \midrule

XGBoost Classifier & Combined All
& 58.09 & 58.09 & 53.29 & 52.86 & \textbf{42.15} \\
& Comments Only
& 58.09 & 58.09 & 55.38 & 55.16 & 6.53 \\
& Dead Code Only
& 58.09 & 58.09 & 56.89 & 56.28 & 29.70 \\
& Structure Only
& 58.09 & 58.09 & 56.76 & 56.40 & 28.15 \\ \midrule

Random Forest & Combined All
& 55.38 & 55.37 & 52.53 & 52.41 & \textbf{40.92} \\
(100 Trees) & Comments Only
& 55.38 & 55.37 & 54.71 & 54.67 & 12.01 \\
& Dead Code Only
& 55.38 & 55.37 & 55.24 & 55.20 & 25.12 \\
& Structure Only
& 55.38 & 55.37 & 54.00 & 53.94 & 30.02 \\ \midrule

k-NN ($k=5$) & Combined All
& 52.76 & 52.75 & 52.31 & 52.29 & \textbf{37.63} \\
& Comments Only
& 52.76 & 52.75 & 52.53 & 52.51 & 12.88 \\
& Dead Code Only
& 52.76 & 52.75 & 52.62 & 52.61 & 28.60 \\
& Structure Only
& 52.76 & 52.75 & 52.98 & 52.98 & 29.93 \\

\bottomrule
\end{tabular}
\end{table*}

\smallskip
\noindent\textbf{Step 2: Deterministic Mutation Generation}

For clean function $x_i$, each mutation track is defined as
\vspace{-0.5em}
\begin{equation}
x_i^{(t)}=m_t(x_i),
\qquad
t\in\{c,d,s,a\},
\label{eq:mutation}
\end{equation}

where $c$, $d$, $s$, and $a$ denote comment removal, dead-code
insertion, structural rewriting, and their composition, respectively.

\noindent\textbf{1. Comments Only:} Tree-sitter comment nodes are replaced
with whitespace while preserving newline characters. Comment-like
sequences inside strings are not modified.

\noindent\textbf{2. Dead Code Only:} A deterministic unreachable conditional
block is inserted immediately after the opening brace of the first detected
function body. Generated variable names are checked against identifiers
already present in the function.

\noindent\textbf{3. Structure Only:} Conservatively selected parameters and
local variables are renamed using deterministic pair-derived identifiers.
Eligible \texttt{for} loops are converted into scoped \texttt{while} loops.
Potentially shadowed variables, incomplete loops, and loops containing a
\texttt{continue} statement that would alter update behaviour are skipped.

\noindent\textbf{4. Combined All:} Comment removal, dead-code insertion,
identifier renaming, and loop conversion are applied sequentially.

The C and C++ Tree-sitter parsers are evaluated when available, and the
parser producing the fewest explicit error or missing nodes is selected.
A candidate mutation is rejected if it produces empty code or increases
the number of parse issues relative to the original function. This
validation prevents parse regression but does not provide a formal proof
of compilation or semantic equivalence.

For each sample and track, the implementation records whether the code
changed, whether the transformation was applied, its status, parse-issue
count, error information, and SHA-256 hash. When no applicable target
exists, the clean function is retained and recorded as a valid unchanged
case.

\begin{table*}[t]
\centering
\small
\caption{Test-Set Post-Defense Multi-Track Mutation Evaluation Matrix for the Validation-Selected 256-Dimensional Representation. Reported values are means across the three projector seeds, with standard deviations reported in the accompanying results text or supplementary tables.}
\label{tab:defended_matrix}
\begin{tabular}{llccccc}
\toprule
\textbf{Classifier Model} & \textbf{Mutation Track} &
\textbf{C-Acc (\%)} & \textbf{C-F1 (\%)} &
\textbf{M-Acc (\%)} & \textbf{M-F1 (\%)} &
\textbf{VFR (\%)} \\ \midrule

MLP Classifier & Combined All
& 58.01 & 55.40 & 56.70 & 52.55 & \textbf{5.89} \\
(128, 64) & Comments Only
& 58.01 & 55.40 & 56.74 & 52.35 & 1.35 \\
& Dead Code Only
& 58.01 & 55.40 & 58.36 & 56.20 & 4.73 \\
& Structure Only
& 58.01 & 55.40 & 57.36 & 54.50 & 4.97 \\ \midrule

Logistic Regression & Combined All
& 59.39 & 59.36 & 56.84 & 56.23 & \textbf{17.50} \\
& Comments Only
& 59.39 & 59.36 & 57.97 & 57.55 & 3.54 \\
& Dead Code Only
& 59.39 & 59.36 & 58.71 & 58.64 & 13.68 \\
& Structure Only
& 59.39 & 59.36 & 58.33 & 58.00 & 10.16 \\ \midrule

SVM (RBF Kernel) & Combined All
& 58.73 & 57.54 & 56.96 & 53.92 & \textbf{4.63} \\
& Comments Only
& 58.73 & 57.54 & 57.54 & 54.85 & 1.06 \\
& Dead Code Only
& 58.73 & 57.54 & 58.53 & 57.53 & 5.66 \\
& Structure Only
& 58.73 & 57.54 & 57.97 & 56.16 & 3.38 \\ \midrule

XGBoost Classifier & Combined All
& 58.47 & 58.40 & 57.19 & 56.52 & \textbf{14.36} \\
& Comments Only
& 58.47 & 58.40 & 57.26 & 56.67 & 4.12 \\
& Dead Code Only
& 58.47 & 58.40 & 57.97 & 57.95 & 13.44 \\
& Structure Only
& 58.47 & 58.40 & 58.43 & 58.05 & 9.81 \\ \midrule

Random Forest & Combined All
& 57.50 & 57.49 & 55.91 & 55.83 & \textbf{28.51} \\
(100 Trees) & Comments Only
& 57.50 & 57.49 & 57.04 & 56.92 & 10.10 \\
& Dead Code Only
& 57.50 & 57.49 & 58.06 & 58.05 & 19.73 \\
& Structure Only
& 57.50 & 57.49 & 56.74 & 56.72 & 21.86 \\ \midrule

k-NN ($k=5$) & Combined All
& 55.32 & 55.32 & 54.27 & 54.25 & \textbf{35.27} \\
& Comments Only
& 55.32 & 55.32 & 55.36 & 55.34 & 14.29 \\
& Dead Code Only
& 55.32 & 55.32 & 55.42 & 55.41 & 28.34 \\
& Structure Only
& 55.32 & 55.32 & 54.73 & 54.72 & 28.12 \\

\bottomrule
\end{tabular}
\end{table*}

\smallskip
\noindent\textbf{Step 3: Dataset Splitting}

The complete paired and mutated benchmark is divided into training,
validation, and test sets using target ratios of 70\%, 15\%, and 15\%.
The realized sizes may differ slightly because complete groups cannot
be divided across splits.

A union--find procedure first connects samples sharing the same
\texttt{split\_group\_id} and ensures that both members of each
vulnerable--patched pair remain together. SHA-256 hashes are then
computed for the clean and all mutated code tracks. Groups containing
identical code in any track are merged, including cases where one
sample's mutated code is identical to another sample's clean code.

Complete effective groups are assigned using a randomized greedy
procedure that approximates the requested split proportions and class
balance. Five hundred deterministic candidate assignments are evaluated
using random seed 42, and the assignment with the lowest imbalance
score is retained. A final audit verifies that no
\texttt{sample\_id}, \texttt{pair\_id}, effective group, or exact code
hash crosses the dataset splits.

The training split is used to optimize the projection networks and fit
the downstream classifiers. The validation split is used for projector
early stopping, dimensional-ablation selection, and checkpoint
selection. The test split remains untouched until the latent dimension
and all other configurations have been fixed.

\smallskip
\noindent\textbf{Step 4: Frozen CodeBERT Feature Extraction}

All clean and mutated functions are embedded using
\texttt{microsoft/codebert-base} with the fast Hugging Face tokenizer.
Inputs are padded within each batch and truncated to a maximum of 512
tokens, including special tokens. For sample $i$ and mutation track $t$,
the frozen representation is

\begin{equation}
h_i^{(t)}
=
E_{\mathrm{CB}}\!\left(x_i^{(t)}\right)_{[\mathrm{CLS}]}
\in\mathbb{R}^{768},
\label{eq:embedding}
\end{equation}

where $E_{\mathrm{CB}}$ denotes CodeBERT and the first-token
representation from the final hidden layer is used. Embeddings are
extracted in batches of 16 as 32-bit floating-point vectors. CodeBERT
remains frozen and is placed in evaluation mode. Extraction is performed
under \texttt{torch.inference\_mode()}, which disables dropout, gradient
tracking, and parameter updates. Python, NumPy, and PyTorch use random
seed 42, and deterministic CUDA behaviour is requested when a GPU is
available. Repeated extraction of the same input is checked for
determinism. The pipeline rejects non-finite and near-zero embeddings and
stores the full token count, used token count, and truncation indicator
for every sample and mutation track.

\smallskip
\noindent\textbf{Step 5: Baseline Downstream Evaluation}

Six downstream classifier families are evaluated using random seed 42:

\noindent\textbf{1. Logistic Regression:} Standardized inputs and a maximum
of 2,000 optimization iterations.

\noindent\textbf{2. RBF SVM:} Standardized inputs and an RBF kernel.

\noindent\textbf{3. Random Forest:} 100 trees.

\noindent\textbf{4. XGBoost:} 100 estimators, a learning rate of 0.05,
and a maximum tree depth of 6.

\noindent\textbf{5. $k$-NN:} Standardized inputs and $k=5$.

\noindent\textbf{6. MLP:} Standardized inputs, hidden layers of sizes 128
and 64, a maximum of 500 iterations, and early stopping.

Parameters not explicitly listed retain their corresponding library
defaults. Logistic Regression, SVM, $k$-NN, and MLP use pipelines in
which the scaler is fit only on clean training embeddings. Random Forest
and XGBoost operate directly on the unstandardized representations.

Each classifier is trained exclusively on clean training embeddings and
then evaluated on the clean and four mutated views of the validation and
test sets. Mutation embeddings are not used to train the downstream
classifiers.

All six classifier families are retained in the final robustness and
statistical evaluation. This avoids basing the conclusions on a single
decision-boundary geometry and allows the analysis to distinguish
representation-level improvements from classifier-specific effects.
Logistic Regression may additionally be presented as a compact
representative case, but the final baseline--defence comparisons are
performed for every classifier.

\smallskip
\noindent\textbf{Step 6: Joint Classification--Invariance Defence and
Dimensional Ablation}

A learned projector maps each frozen CodeBERT embedding through a
two-layer network,

\begin{equation}
P_d:
\mathbb{R}^{768}
\rightarrow
\mathbb{R}^{512}
\rightarrow
\mathbb{R}^{d},
\qquad
d\in\{128,256,512\}.
\label{eq:projector}
\end{equation}

The first linear layer is followed by LayerNorm, ReLU, and dropout with
probability 0.1. The final $d$-dimensional vector is L2-normalized to
produce the defended representation $z_i^{(t)}$. A two-class auxiliary
linear classifier is used only while training the projector.

For a batch of size $B$, let $\mathrm{CE}_0$ denote cross-entropy on
clean embeddings and let $\mathrm{CE}_{\mathrm{mut}}$ denote
cross-entropy averaged across the four mutation tracks. The training
objective is

\vspace{-1em}
\begin{equation}
\begin{aligned}
\mathcal{L}_{\mathrm{joint}}
&=
\mathcal{L}_{\mathrm{cls}}
+\alpha\mathcal{L}_{\mathrm{inv}},\\
\mathcal{L}_{\mathrm{cls}}
&=
\tfrac{1}{2}
\left(
\mathrm{CE}_0+\mathrm{CE}_{\mathrm{mut}}
\right),\\
\mathcal{L}_{\mathrm{inv}}
&=
\tfrac{1}{4B}
\sum_{i=1}^{B}
\sum_{t\in\{c,d,s,a\}}
\left(
1-z_i^{(0)\top}z_i^{(t)}
\right),
\end{aligned}
\label{eq:joint_loss}
\end{equation}

with $\alpha=1.0$. Because the defended representations are
L2-normalized, their inner product corresponds to cosine similarity.

The projector is trained using AdamW with learning rate
$5\times10^{-4}$, weight decay $10^{-3}$, batch size 128, and at most
100 epochs. Gradients are clipped to a maximum norm of 5.0. A
\texttt{ReduceLROnPlateau} scheduler halves the learning rate after three
validation epochs without improvement. Training stops after ten epochs
without an improvement of at least $10^{-5}$ in validation joint loss,
and the checkpoint with the lowest validation loss is restored.

Each dimensional configuration is trained independently using projector
seeds 42, 43, and 44. The downstream classifier configurations and their
random seeds remain fixed across projector runs so that the observed
variation primarily reflects the learned projection. For every projector
seed and latent dimension, the same six downstream classifier families
are retrained using only clean defended training embeddings and evaluated
on the clean and mutated validation views.

\begin{table}[t]
\centering
\caption{Projection-Dimension Ablation Design}
\label{tab:defense_dimension_ablation}
\footnotesize
\setlength{\tabcolsep}{10pt}
\renewcommand{\arraystretch}{1.05}

\begin{tabular}{@{}ccc@{}}
\toprule
\shortstack{\textbf{Latent}\\\textbf{Dimension}} &
\shortstack{\textbf{Projection}\\\textbf{Architecture}} &
\shortstack{\textbf{Projector}\\\textbf{Seeds}} \\
\midrule
128 & $768 \rightarrow 512 \rightarrow 128$ & 42, 43, 44 \\
256 & $768 \rightarrow 512 \rightarrow 256$ & 42, 43, 44 \\
512 & $768 \rightarrow 512 \rightarrow 512$ & 42, 43, 44 \\
\bottomrule
\end{tabular}
\end{table}

Architecture selection is performed using validation data only. For
each latent dimension, metrics are first summarized across the six
classifiers and four mutation tracks for each projector seed and then
reported as means and standard deviations across the three seeds. A
configuration is eligible only when its mean clean validation macro-F1
is no more than one percentage point below the corresponding baseline
mean. Among eligible configurations, the selected dimension is the one
with the lowest mean validation VFR. Remaining ties are resolved using,
in order, the highest mean mutated macro-F1, the lowest mean
benign-to-vulnerable flip rate, and the smaller latent dimension.

Under this predefined rule, the 256-dimensional configuration is
selected. This selection indicates that it achieved the best validation
criterion among the tested configurations; it is not treated as evidence
that 256 dimensions are statistically superior to every alternative.
After selection, the three 256-dimensional checkpoints are applied
unchanged to the test split. Test data are used only for final evaluation
and do not influence the latent dimension, early-stopping epoch, or
checkpoint selection.

After projection, the same six classifier families are retrained only on
clean defended training embeddings for each selected projector seed. The
auxiliary classifier used during projector learning is not used in the
final classifier evaluation. Since all four mutation families are
included during defence training, the defended test evaluation measures
generalization to unseen samples drawn from known transformation
families rather than to previously unseen mutation operators.

\smallskip
\noindent\textbf{Step 7: Evaluation Metrics}

For classifier $f$ and representation $r_i^{(t)}$, where
$r_i^{(t)}$ denotes either a baseline or defended embedding, the
predicted label is

\begin{equation}
\hat{y}_i^{(t)}=f\!\left(r_i^{(t)}\right).
\end{equation}

Clean and mutated accuracy and macro-F1 are reported for every
classifier and mutation track. Overall prediction flip rate is the
percentage of samples whose predicted label changes between the clean
and mutated inputs.

The main class-conditional metric is Vulnerable Flip Rate (VFR), which
measures the percentage of clean true-positive vulnerable samples that
become benign predictions after mutation:

\vspace{-1em}
\begin{equation}
\mathrm{VFR}^{(t)}
=
\frac{
\left|
\left\{
i:
y_i=1,\,
\hat{y}_i^{(0)}=1,\,
\hat{y}_i^{(t)}=0
\right\}
\right|
}{
\left|
\left\{
i:
y_i=1,\,
\hat{y}_i^{(0)}=1
\right\}
\right|
}
\times100\%.
\label{eq:vfr}
\end{equation}

The reverse benign-to-vulnerable flip rate is calculated analogously
among clean true-negative benign samples. The numerator and denominator
of every conditional metric are retained, and a zero denominator is
reported as undefined rather than as a zero rate. Mutated false-negative
and false-positive rates and the full paired transition counts are also
recorded.

For standalone baseline and defended evaluations, each representation's
VFR denominator is defined by its own clean true-positive samples. For
direct baseline--defence comparisons, common clean subsets are used to
ensure that both models are evaluated with identical denominators.
Common-set VFR is calculated among vulnerable samples that are clean
true positives under both representations. The common-set
benign-to-vulnerable rate is calculated among benign samples that are
clean true negatives under both representations.

\smallskip
\noindent\textbf{Step 8: Statistical Testing and Visualization}

Within each representation, clean and mutated correctness are paired for
the same test samples. Exact two-sided McNemar tests compare
clean-correct-to-mutated-wrong transitions against
clean-wrong-to-mutated-correct transitions. Holm correction is applied
across the four mutation tracks separately for each classifier,
representation, and projector seed.

VFR and benign-to-vulnerable flip-rate confidence intervals are
estimated using 5,000 grouped bootstrap iterations. Complete groups are
sampled with replacement using the strongest available grouping key in
the following order:
\texttt{effective\_split\_group}, \texttt{split\_group\_id},
\texttt{pair\_id}, and \texttt{sample\_id}. The 2.5th and 97.5th
percentiles form the 95\% confidence interval.

A separate paired baseline--defence analysis aligns samples by
\texttt{sample\_id}. Mutated correctness is compared using an exact
two-sided McNemar test. To compare VFR using a common denominator, the
analysis is restricted to vulnerable samples correctly classified in
both the clean baseline and clean defended spaces. Baseline and defended
vulnerable-flip events are then compared using an exact McNemar test, and
the baseline-minus-defended VFR reduction is accompanied by a grouped
bootstrap confidence interval.

The false-positive trade-off is evaluated analogously using benign
samples correctly classified in both clean representation spaces.
Baseline and defended benign-to-vulnerable flip events are compared
using an exact McNemar test, and the defended-minus-baseline rate change
is accompanied by a grouped bootstrap confidence interval.

Holm correction is applied separately across the four mutation tracks
for each classifier, projector seed, and statistical family: mutated
accuracy, common-set VFR, and common-set benign-flip rate. Statistical
tests are therefore not pooled across classifiers, projector seeds, or
outcome families.

Final defended performance and effect sizes are summarized
descriptively as means and standard deviations across projector seeds
42, 43, and 44. The three projector runs are evaluated on the same
held-out test samples and are not treated as independent copies of the
test set. Their predictions are not pooled into a larger significance
test, and their $p$-values are not averaged. Statistical significance is
reported separately for each seed, while cross-seed summaries report the
number of seeds for which the corrected test is significant.

\begin{table*}[t]
\centering
\footnotesize
\caption{Paired test-set comparison between the baseline and the
validation-selected 256-dimensional defence. Effect sizes are reported
as mean $\pm$ standard deviation across three projector seeds.}
\label{tab:statistical_tests}
\setlength{\tabcolsep}{4pt}
\renewcommand{\arraystretch}{1.08}

\begin{tabular}{@{}llcccc@{}}
\toprule
\textbf{Classifier} &
\textbf{Mutation Track} &
\shortstack{\textbf{$\Delta$Acc}\\\textbf{(pp)}} &
\shortstack{\textbf{VFR Reduction}\\\textbf{(pp)}} &
\shortstack{\textbf{VFR Sig.}\\\textbf{Seeds}} &
\shortstack{\textbf{B$\rightarrow$V Change}\\\textbf{(pp)}} \\
\midrule

Logistic Regression
& Combined All
& $0.49 \pm 0.42$
& $17.29 \pm 9.37$
& \textbf{3/3}
& $11.95 \pm 8.72$ \\

& Comments Only
& $0.41 \pm 0.27$
& $2.13 \pm 0.55$
& 1/3
& $6.52 \pm 1.39$ \\

& Dead Code Only
& $1.33 \pm 0.73$
& $6.76 \pm 3.41$
& 2/3
& $-0.14 \pm 1.85$ \\

& Structure Only
& $1.17 \pm 0.65$
& $13.25 \pm 5.42$
& \textbf{3/3}
& $8.89 \pm 6.00$ \\
\midrule

SVM (RBF Kernel)
& Combined All
& $0.43 \pm 0.09$
& $34.73 \pm 1.44$
& \textbf{3/3}
& $19.69 \pm 3.11$ \\

& Comments Only
& $-0.33 \pm 0.46$
& $2.03 \pm 0.22$
& \textbf{3/3}
& $3.67 \pm 1.09$ \\

& Dead Code Only
& $0.00 \pm 0.28$
& $17.44 \pm 0.67$
& \textbf{3/3}
& $3.65 \pm 1.24$ \\

& Structure Only
& $-0.61 \pm 0.36$
& $16.21 \pm 0.45$
& \textbf{3/3}
& $10.90 \pm 1.83$ \\
\midrule

Random Forest
& Combined All
& $3.33 \pm 0.80$
& $12.20 \pm 5.78$
& \textbf{3/3}
& $0.81 \pm 2.54$ \\

(100 Trees)
& Comments Only
& $2.33 \pm 1.18$
& $1.43 \pm 1.06$
& 0/3
& $-0.34 \pm 1.97$ \\

& Dead Code Only
& $2.81 \pm 0.33$
& $6.25 \pm 1.19$
& 2/3
& $0.99 \pm 2.35$ \\

& Structure Only
& $2.70 \pm 0.88$
& $4.59 \pm 2.56$
& 2/3
& $-1.04 \pm 1.64$ \\
\midrule

XGBoost
& Combined All
& $3.90 \pm 0.63$
& $25.95 \pm 3.78$
& \textbf{3/3}
& $8.99 \pm 5.76$ \\

& Comments Only
& $1.88 \pm 0.81$
& $3.56 \pm 0.79$
& \textbf{3/3}
& $0.55 \pm 0.58$ \\

& Dead Code Only
& $1.08 \pm 0.53$
& $14.76 \pm 2.74$
& \textbf{3/3}
& $2.63 \pm 2.03$ \\

& Structure Only
& $1.67 \pm 0.40$
& $14.37 \pm 2.36$
& \textbf{3/3}
& $7.24 \pm 2.05$ \\
\midrule

k-NN ($k=5$)
& Combined All
& $1.96 \pm 0.73$
& $4.19 \pm 2.23$
& 0/3
& $-2.34 \pm 2.67$ \\

& Comments Only
& $2.83 \pm 1.09$
& $-1.58 \pm 0.86$
& 0/3
& $-1.06 \pm 2.38$ \\

& Dead Code Only
& $2.80 \pm 0.44$
& $1.40 \pm 1.22$
& 0/3
& $3.30 \pm 3.83$ \\

& Structure Only
& $1.75 \pm 0.90$
& $4.01 \pm 3.48$
& 0/3
& $-0.76 \pm 2.37$ \\
\midrule

MLP (128, 64)
& Combined All
& $2.83 \pm 0.51$
& $30.45 \pm 2.05$
& \textbf{3/3}
& $20.16 \pm 2.78$ \\

& Comments Only
& $1.76 \pm 0.44$
& $2.60 \pm 0.19$
& \textbf{3/3}
& $14.15 \pm 1.28$ \\

& Dead Code Only
& $4.18 \pm 0.40$
& $22.67 \pm 1.45$
& \textbf{3/3}
& $-1.19 \pm 1.24$ \\

& Structure Only
& $1.76 \pm 0.07$
& $14.13 \pm 1.27$
& \textbf{3/3}
& $7.90 \pm 1.33$ \\

\bottomrule
\end{tabular}

\vspace{1mm}
\begin{minipage}{0.98\textwidth}
\footnotesize
\textit{Notes:}
$\Delta$Acc is defended minus baseline mutated accuracy.
VFR Reduction is baseline minus defended common-set Vulnerable Flip
Rate; positive values indicate improved vulnerability preservation.
VFR Sig.\ Seeds reports the number of projector seeds with a
Holm-adjusted exact McNemar $p<0.05$.
B$\rightarrow$V Change is defended minus baseline common-set
benign-to-vulnerable flip rate; positive values indicate an increase in
false-positive transitions. Complete per-seed statistical tests and
grouped-bootstrap confidence intervals are reported separately.
\end{minipage}
\end{table*}

PCA and UMAP reducers are fit only on training embeddings and then
applied to held-out test embeddings from the corresponding representation
and projector seed. These two-dimensional projections are treated as
qualitative summaries rather than direct evidence of classification
performance. When a single projector seed is displayed, it is identified
explicitly and is not used as a substitute for the three-seed aggregate
evaluation.

Embedding displacement is summarized using cosine distance,

\vspace{-1em}
\begin{equation}
d_{\cos}(r,r')=1-\cos(r,r').
\end{equation}

Cosine distance is used for comparisons involving spaces with different
dimensions and scales. Euclidean distance is interpreted only within a
single representation space. Sample-level displacement figures are
illustrative and are not used in aggregate metrics or statistical
tests.

\section{Experiments and Results}

\smallskip
\noindent\textbf{Baseline Global Space Topology}

Figure~\ref{fig:baseline_global_separation} shows PCA and UMAP
projections of the held-out clean test embeddings after fitting each
reducer only on the baseline training embeddings. Benign and vulnerable
samples remain substantially overlapped in both two-dimensional views.
This observation is consistent with the modest clean test performance of
the downstream classifiers, although the projections alone cannot
establish separability in the complete 768-dimensional representation
space.

\smallskip
\noindent\textbf{Baseline Multi-Track Classifier Results}

Table~\ref{tab:ablation_matrix} reports final test performance for all
six baseline classifiers. The RBF SVM obtains the highest clean test
accuracy and macro-F1, at 59.42\% and 59.23\%, respectively. Combined
mutations produce the highest VFR for every classifier, ranging from
35.55\% for Logistic Regression to 42.15\% for XGBoost. Comments Only
is the least disruptive isolated track, with VFR values between 3.22\%
and 12.88\%. Dead-code and structural transformations produce
substantially larger vulnerable-to-benign flip rates, demonstrating that
clean accuracy alone does not characterize robustness under non-functional
source-code changes.

\smallskip
\noindent\textbf{Projection-Dimension Ablation}

Table~\ref{tab:defense_dimension_results} summarizes the validation-only
ablation over latent dimensions 128, 256, and 512. Each configuration is
evaluated using the same six classifier families, four mutation tracks,
and projector seeds 42, 43, and 44. All three configurations satisfy the
predefined eligibility condition because their mean clean validation
macro-F1 remains within one percentage point of the baseline mean of
58.17\%. The 256-dimensional projector obtains the lowest mean validation VFR at
$11.03\%\pm1.49\%$, compared with $11.20\%\pm0.91\%$ for 128
dimensions and $13.12\%\pm0.90\%$ for 512 dimensions. It is therefore
selected by the predefined robustness-first validation rule. However,
the 128- and 512-dimensional configurations obtain higher clean and
mutated macro-F1 values. Moreover, the VFR difference between 128 and
256 dimensions is only 0.17 percentage points. The selection should
therefore be interpreted as the outcome of the specified validation
criterion rather than evidence that 256 dimensions are statistically or
universally superior to the alternatives.

\begin{table*}[t]
\centering
\small
\caption{Validation-set projection-dimension ablation. Values are
reported as mean $\pm$ standard deviation across projector seeds 42,
43, and 44 after averaging across six classifiers and four mutation
tracks.}
\label{tab:defense_dimension_results}
\setlength{\tabcolsep}{6pt}
\renewcommand{\arraystretch}{1.08}
\begin{tabular}{cccccc}
\toprule
\textbf{Latent Dim.} &
\textbf{Clean F1 (\%)} &
\textbf{Mutated F1 (\%)} &
\textbf{VFR (\%)} &
\textbf{B$\rightarrow$V (\%)} &
\textbf{Status} \\
\midrule

128 &
$59.08 \pm 0.61$ &
$58.24 \pm 0.42$ &
$11.20 \pm 0.91$ &
$20.70 \pm 0.69$ &
Eligible \\

\textbf{256} &
$58.29 \pm 0.48$ &
$57.60 \pm 0.05$ &
$\mathbf{11.03 \pm 1.49}$ &
$20.45 \pm 1.39$ &
\textbf{Selected} \\

512 &
$59.13 \pm 0.12$ &
$58.29 \pm 0.27$ &
$13.12 \pm 0.90$ &
$\mathbf{19.33 \pm 1.11}$ &
Eligible \\

\bottomrule
\end{tabular}
\end{table*}

\smallskip
\noindent\textbf{Defended Global Space Topology}

Figure~\ref{fig:defended_global_separation} presents PCA and UMAP
projections after freezing the validation-selected projector. The
displayed representation uses projector seed 44, which is the
median-validation-VFR run among the three selected 256-dimensional
checkpoints. Each dimensionality reducer is fitted only on the defended
training embeddings and is then applied to the corresponding held-out
test embeddings. The projection changes both local and global neighbourhood geometry, but
the two classes remain partially overlapped in the two-dimensional
views. These visualizations therefore do not establish complete class
separation or prediction invariance. The classifier metrics and paired
prediction-transition analyses provide the primary quantitative evidence.

\begin{figure*}[t]
\centering
\includegraphics[width=\textwidth]
{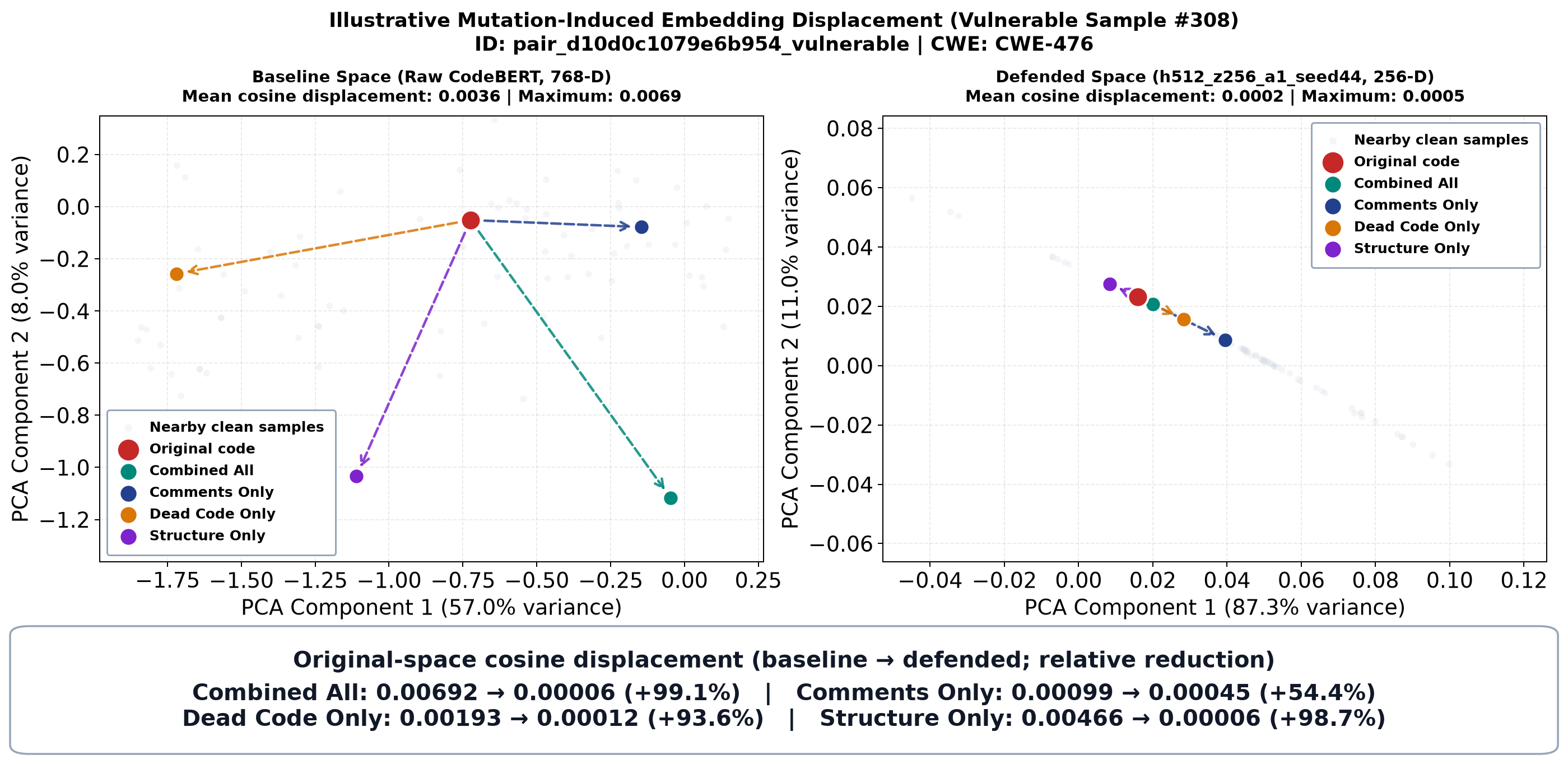}
\caption{Illustrative mutation-induced embedding displacement for one
vulnerable held-out test sample. The left panel displays the clean
baseline embedding and all four mutation tracks in the
768-dimensional CodeBERT space. The right panel displays the
corresponding points in the validation-selected 256-dimensional defended
space for projector seed 44. Separate PCA reducers are fitted on the
corresponding baseline and defended training embeddings. PCA coordinates
and arrow geometry are qualitative. Numerical comparisons below the
panels use clean-to-mutated cosine distance in the original
representation spaces.}
\label{fig:sample_vector_displacement}
\end{figure*}

\smallskip
\noindent\textbf{Post-Defense Multi-Track Classifier Results}

Table~\ref{tab:defended_matrix} reports final defended test performance
as means across projector seeds 42, 43, and 44. Averaged across the six
classifier families and four mutation tracks, the selected defence
obtains clean accuracy of $57.90\%\pm0.36\%$, clean macro-F1 of
$57.25\%\pm0.32\%$, mutated accuracy of $57.10\%\pm0.37\%$, and
mutated macro-F1 of $56.04\%\pm0.27\%$. Its overall prediction flip
rate is $17.56\%\pm0.37\%$. The standalone defended VFR is
$12.69\%\pm1.93\%$, while the benign-to-vulnerable flip rate is
$21.43\%\pm1.61\%$.

For Combined All mutations, standalone VFR decreases descriptively for
every classifier relative to the corresponding baseline result. The
smallest remaining combined-track VFR values occur for the SVM and MLP,
at 4.63\% and 5.89\%, respectively. XGBoost records 14.36\%,
Logistic Regression 17.50\%, Random Forest 28.51\%, and $k$-NN
35.27\%. The learned projection therefore interacts differently with
the downstream decision boundaries. Its strongest standalone results
occur for the SVM, MLP, and XGBoost, whereas $k$-NN retains substantial
sensitivity to the combined transformation. The standalone rates in Table~\ref{tab:defended_matrix} use each
model's own clean true-positive denominator. They characterize the final
defended classifiers but do not provide an equal-denominator
baseline-to-defence comparison. Direct improvement is therefore
evaluated using aligned samples and common clean subsets.

\smallskip
\noindent\textbf{Paired Baseline--Defence Statistical Comparison}

Table~\ref{tab:statistical_tests} reports paired baseline-versus-defence
effects for every classifier and mutation track. Inferential tests are
performed separately for projector seeds 42, 43, and 44. Effect sizes
are then summarized descriptively across the three seeds. Since the
projector runs use the same held-out test samples, their predictions are
not pooled as independent observations, and their $p$-values are not
averaged. Across all 72 classifier--track--seed comparisons, mutated accuracy increases from 55.33\% under the baseline representation to 57.10\% after projection. This is an average gain of 1.77 percentage points. Mutated accuracy improves in 66 of the 72 comparisons, although only 12 improvements remain significant after Holm correction. The accuracy effect is therefore directionally consistent but generally smaller than the class-conditional VFR effect.

On the common clean true-positive subset, VFR decreases from 22.60\%
to 11.32\%. This is an 11.28-percentage-point reduction and corresponds
to a 49.9\% relative reduction in vulnerable-to-benign flips. VFR
decreases in 67 of the 72 comparisons and is significant after Holm
correction in 52. All 52 significant reductions are also accompanied by
grouped-bootstrap confidence intervals whose lower bounds exceed zero. The VFR reduction is significant in all 12 seed-track comparisons for
the SVM, MLP, and XGBoost classifiers. Logistic Regression is
significant in 9 of 12 comparisons, Random Forest in 7 of 12, and
$k$-NN in none. The defence therefore provides strong evidence of
improved vulnerability preservation for several classifier families,
but the improvement does not generalize equally to every downstream
decision rule.

The largest track-level benefit occurs for Combined All mutations.
Averaged across classifiers and seeds, common-set VFR decreases from
37.36\% to 16.56\%, a reduction of 20.80 percentage points. The
reduction is significant in 15 of the 18 combined-track comparisons.
Dead Code Only and Structure Only also show substantial mean reductions
of 11.55 and 11.09 percentage points, respectively. Comments Only shows
a smaller reduction of 1.69 percentage points because its baseline VFR
is already comparatively low. 
The improved vulnerable-prediction retention is accompanied by a
reverse-direction trade-off. Across all paired comparisons, common-set
benign-to-vulnerable flip rate increases from 14.40\% to 19.62\%, a
mean increase of 5.21 percentage points. The increase occurs in 54 of
72 comparisons and is significant after Holm correction in 32. Thus,
the projection improves directional security robustness by preventing
many vulnerable predictions from becoming benign, but it does not
produce uniformly bidirectional prediction stability and can increase
false-positive transitions.

\smallskip
\noindent\textbf{Granular Sample Vector Displacement}

Figure~\ref{fig:sample_vector_displacement} illustrates how the four
mutation tracks move one vulnerable test sample before and after
projection. For the displayed example, the defended cosine distances
are smaller than their baseline counterparts, although the reduction
varies across mutation types. The two PCA panels use separately fitted
reducers; consequently, their coordinate scales and displayed arrow
lengths must not be compared directly. A mutation point may overlap the clean anchor when the corresponding
transformation does not change the input or produces an effectively
identical representation. For example, comment removal may leave a
function unchanged when it contains no removable comments. The figure
is illustrative and is not used to estimate average performance or
statistical significance.

\smallskip
\noindent\textbf{Qualitative Case Analysis}

To complement the aggregate evaluation, five changed test mutations are
selected using the baseline Logistic Regression model and the seed-44
defended Logistic Regression model. The selected examples satisfy the
available parse-validity checks, and none of their clean or mutated inputs
is truncated by CodeBERT's 512-token limit. They represent a corrected
missed-vulnerability transition, a new missed-vulnerability transition
introduced by the defence, a new false-positive transition introduced by
the defence, and two baseline failures that remain after projection.
These individual cases illustrate possible behaviours and are not used
to estimate their prevalence.

\paragraph{Case 1: Vulnerable-to-Benign Failure Corrected by the Defence.}

The first case is the vulnerable
\texttt{cJSON\_GetObjectItem} function from sample
\texttt{pair\_03866928959097eb\_vulnerable}. Its clean version is:

\begin{lstlisting}[
language=C,
basicstyle=\ttfamily\scriptsize,
breaklines=true,
frame=single,
caption={Clean version of the vulnerable
\texttt{cJSON\_GetObjectItem} sample.},
label={lst:qualitative_corrected_clean}
]
cJSON *cJSON_GetObjectItem(
    cJSON *object,
    const char *string)
{
    cJSON *c = object->child;

    while (c &&
           cJSON_strcasecmp(c->string, string))
        c = c->next;

    /* Remaining statements omitted for space. */
}
\end{lstlisting}

The Combined All mutation renames local identifiers and inserts an
unreachable block near the beginning of the function:

\begin{lstlisting}[
language=C,
basicstyle=\ttfamily\scriptsize,
breaklines=true,
frame=single,
caption={Abridged Combined All mutation of
\texttt{cJSON\_GetObjectItem}.},
label={lst:qualitative_corrected_mutated}
]
cJSON *cJSON_GetObjectItem(
    cJSON *mut_var_6f5daae3ab,
    const char *mut_var_c3b3d91737)
{
    if (((unsigned long)24 *
         (unsigned long)13) == 0UL) {
        unsigned int mut_var_912c7b3763 =
            24u ^ 13u;
    }

    /* Renamed and structurally transformed
       statements omitted for space. */
}
\end{lstlisting}

The clean and mutated functions contain 72 and 224 tokens,
respectively, and neither is truncated. The Combined All mutation causes
the baseline vulnerable-class probability to decrease from 0.890 to
0.482. The baseline prediction therefore changes from vulnerable to
benign. In the defended space, the corresponding probabilities are
0.922 and 0.878, so the defended model retains the vulnerable prediction.

The baseline probability decreases by 0.408, whereas the defended
probability decreases by only 0.044. This case demonstrates the intended
effect of the learned projection: although the mutation still changes
the model score, the transformed sample remains on the vulnerable side
of the defended classifier boundary.

\paragraph{Case 2: New Vulnerable-to-Benign Error Introduced by the Defence.}

The second case is the vulnerable \texttt{user\_match} function from
sample \texttt{pair\_fa10c84b69ae46a1\_vulnerable}. Its clean version is:

\begin{lstlisting}[
language=C,
basicstyle=\ttfamily\scriptsize,
breaklines=true,
frame=single,
caption={Clean version of the vulnerable
\texttt{user\_match} sample.},
label={lst:qualitative_new_vfr_clean}
]
int user_match(
    const struct key *key,
    const struct key_match_data *match_data)
{
    return strcmp(
        key->description,
        match_data->raw_data) == 0;
}
\end{lstlisting}

The Dead Code Only transformation inserts an unreachable block while
leaving the original comparison unchanged:

\begin{lstlisting}[
language=C,
basicstyle=\ttfamily\scriptsize,
breaklines=true,
frame=single,
caption={Dead Code Only mutation of
\texttt{user\_match}.},
label={lst:qualitative_new_vfr_mutated}
]
int user_match(
    const struct key *key,
    const struct key_match_data *match_data)
{
    if (((unsigned long)50 *
         (unsigned long)3) == 0UL) {
        unsigned int mut_dead_25f115594f =
            50u ^ 3u;
    }

    return strcmp(
        key->description,
        match_data->raw_data) == 0;
}
\end{lstlisting}

The clean function contains 51 tokens, while the mutated version contains
106 tokens; neither is truncated. The baseline vulnerable-class
probability decreases from 0.961 to 0.594, but the baseline model
preserves the correct vulnerable prediction. In the defended space, the
probability decreases from 0.995 to 0.408, causing the mutated sample to
be classified as benign.

The baseline probability drop is 0.367, compared with 0.587 after
projection. The inserted block is designed not to execute, but it changes
the lexical and positional content supplied to the encoder. For this
sample, the learned projection amplifies rather than suppresses the
resulting boundary movement, creating a new missed-vulnerability error.

\paragraph{Case 3: New Benign-to-Vulnerable Error Introduced by the Defence.}

The third case is the benign \texttt{cliRefreshPrompt} function from
sample \texttt{pair\_5af43057f57ac4f3\_benign}. An abridged clean
version is:

\begin{lstlisting}[
language=C,
basicstyle=\ttfamily\scriptsize,
breaklines=true,
frame=single,
caption={Abridged clean version of the benign
\texttt{cliRefreshPrompt} sample.},
label={lst:qualitative_new_benign_flip_clean}
]
static void cliRefreshPrompt(void)
{
    if (config.eval_ldb)
        return;

    sds prompt = sdsempty();

    if (config.hostsocket != NULL) {
        prompt = sdscatfmt(
            prompt,
            "redis %s",
            config.hostsocket);
    }

    /* Remaining statements omitted for space. */
}
\end{lstlisting}

The Combined All mutation inserts an unreachable block and renames
eligible local identifiers:

\begin{lstlisting}[
language=C,
basicstyle=\ttfamily\scriptsize,
breaklines=true,
frame=single,
caption={Abridged Combined All mutation of
\texttt{cliRefreshPrompt}.},
label={lst:qualitative_new_benign_flip_mutated}
]
static void cliRefreshPrompt(void)
{
    if (((unsigned long)83 *
         (unsigned long)14) == 0UL) {
        unsigned int mut_var_752c01ec19 =
            83u ^ 14u;
    }

    if (config.eval_ldb)
        return;

    sds mut_var_1bdf4cddaa = sdsempty();

    if (config.hostsocket != NULL) {
        /* Remaining transformed statements
           omitted for space. */
    }
}
\end{lstlisting}

The clean and mutated inputs contain 297 and 470 tokens, respectively,
and neither is truncated. The baseline vulnerable-class probability
changes only from 0.363 to 0.393. Both values remain below the decision
threshold, so the baseline model preserves the benign prediction. In
the defended space, the probability increases from 0.352 to 0.524,
causing the mutated sample to be classified as vulnerable.

The baseline probability increase is 0.030, while the defended increase
is 0.172. This example illustrates the false-positive trade-off observed
in the aggregate paired analysis. The projection can improve retention
of vulnerable predictions while simultaneously moving some benign
samples across the boundary in the opposite direction.

\paragraph{Case 4: Vulnerable-to-Benign Failure Remaining after Defence.}

The fourth case is the vulnerable \texttt{mark\_object} function from
sample \texttt{pair\_9f29fc85bffac01c\_vulnerable}. Its clean version is:

\begin{lstlisting}[
language=C,
basicstyle=\ttfamily\scriptsize,
breaklines=true,
frame=single,
caption={Clean version of the vulnerable
\texttt{mark\_object} sample.},
label={lst:qualitative_remaining_vfr_clean}
]
static void mark_object(
    struct object *obj,
    struct strbuf *path,
    const char *name,
    void *data)
{
    update_progress(data);
}
\end{lstlisting}

The Combined All mutation renames the parameters and inserts an
unreachable block:

\begin{lstlisting}[
language=C,
basicstyle=\ttfamily\scriptsize,
breaklines=true,
frame=single,
caption={Abridged Combined All mutation of
\texttt{mark\_object}.},
label={lst:qualitative_remaining_vfr_mutated}
]
static void mark_object(
    struct object *mut_var_a31e48846e,
    struct strbuf *mut_var_331eda2ebd,
    const char *mut_var_55aa7a2225,
    void *mut_var_45c0af371d)
{
    if (((unsigned long)29 *
         (unsigned long)19) == 0UL) {
        long mut_var_8c361aecaf =
            (long)29 * (long)19;
    }

    /* Remaining transformed statements
       omitted for space. */
}
\end{lstlisting}

The clean and mutated inputs contain 46 and 148 tokens, respectively,
and neither is truncated. The baseline vulnerable-class probability
decreases from 0.645 to 0.237, causing the prediction to change from
vulnerable to benign. The defended probabilities are 0.539 for the
clean function and 0.484 for the mutated function, so the defended model
also changes from vulnerable to benign.

The projection reduces the probability drop from 0.408 to 0.055.
However, the clean defended sample begins only slightly above the
decision threshold, and the smaller decrease is still sufficient to
cross it. The defence therefore reduces the mutation-induced score
change without correcting this particular prediction failure.

\paragraph{Case 5: Benign-to-Vulnerable Failure Remaining after Defence.}

The fifth case is the benign sample
\texttt{pair\_52144c21c245e11e\_benign}. The relevant clean fragment
contains two explanatory comments surrounding the final assertion:

\begin{lstlisting}[
language=C++,
basicstyle=\ttfamily\scriptsize,
breaklines=true,
frame=single,
caption={Abridged clean version of the benign
Comments Only case.},
label={lst:qualitative_remaining_benign_flip_clean}
]
DiscardReason discard_reason)
{
    DecisionDetails decision_details;

    EXPECT_FALSE(
        lifecycle_unit->CanDiscard(
            discard_reason,
            &decision_details));

    EXPECT_FALSE(
        decision_details.IsPositive());

    // reasons() contains the status for the
    // four local-site feature heuristics,
    // or is empty when no status is tracked.
    EXPECT_TRUE(
        decision_details.reasons().empty() ||
        decision_details.reasons().size() == 4);
}
\end{lstlisting}

The Comments Only transformation removes the explanatory comment while
retaining the surrounding executable statements:

\begin{lstlisting}[
    language=C++,
    basicstyle=\ttfamily\scriptsize,
    breaklines=true,
    frame=single,
    captionpos=t,
    caption={Abridged Comments Only mutation of the benign test sample.},
    label={lst:qualitative_remaining_benign_flip_mutated}
]
DiscardReason discard_reason)
{
    DecisionDetails decision_details;

    EXPECT_FALSE(
        lifecycle_unit->CanDiscard(
            discard_reason,
            &decision_details));

    EXPECT_FALSE(
        decision_details.IsPositive());

    EXPECT_TRUE(
        decision_details.reasons().empty() ||
        decision_details.reasons().size() == 4);
}
\end{lstlisting}

The clean input contains 209 tokens, while the comment-stripped input
contains 169 tokens; neither is truncated. Removing the comments
increases the baseline vulnerable-class probability from 0.239 to 0.533,
causing a benign-to-vulnerable prediction change. The defended model
behaves similarly: its vulnerable-class probability increases from
0.293 to 0.554, and the mutated sample is again classified as vulnerable.

The probability increase is 0.294 in the baseline representation and
0.261 in the defended representation. Because neither input is
truncated, the transition cannot be attributed to CodeBERT's token
limit. Instead, it is consistent with sensitivity to removed comment
tokens or their surrounding lexical context. The case also shows that
the learned projection does not eliminate every pre-existing
mutation-induced false-positive transition.

Overall, the qualitative examples are consistent with the aggregate
findings. The defence can correct harmful vulnerable-to-benign boundary
crossings and can reduce the magnitude of some remaining score changes.
However, it does not provide universal prediction invariance and can
introduce new missed-vulnerability and false-positive errors. These
cases reinforce the need to report both vulnerable-to-benign and
benign-to-vulnerable transitions rather than relying only on clean and
mutated accuracy.

\smallskip
\noindent\textbf{Threats to Validity}

Several limitations affect interpretation. First, patched functions are
treated as benign, although a patch may be incomplete or may contain
unrelated changes. Second, Tree-sitter parse preservation and guarded
rewrites reduce mutation risk but do not constitute formal
semantic-equivalence proofs. Third, CodeBERT inputs longer than 512 tokens
are truncated; truncation is audited but may still influence robustness.
Fourth, the study evaluates one frozen encoder and a single dataset.
Fifth, the defense is trained on the same four mutation families later
used on unseen test samples, so held-out-track experiments are still
required to establish generalization to novel transformations. Sixth, PCA
and UMAP can distort high-dimensional geometry, and the selected
sample-level plot is illustrative. Finally, lower VFR can be achieved
partly by shifting predictions toward the vulnerable class, which is why
benign-to-vulnerable flips and false-positive rates must be reported
alongside VF reductions.

\section{Conclusion}

This study evaluates the robustness of frozen CodeBERT representations
under four deterministic, parse-valid mutation tracks designed to preserve
the intended behaviour of the source code. Combined mutations produce the
largest baseline vulnerability-preservation failures across all six
classifiers, with VFR ranging from 35.55\% to 42.15\%. The
validation-selected joint classification--invariance projector improves
mutated accuracy from 55.33\% to 57.10\% across the paired
baseline--defence comparisons. On the common clean true-positive subset,
VFR decreases from 22.60\% to 11.32\%, an 11.28-percentage-point or
49.9\% relative reduction. The reduction occurs in 67 of 72
classifier--track--seed comparisons and remains significant after Holm
correction in 52.

The improvement is nevertheless classifier- and direction-dependent.
Common-set benign-to-vulnerable flip rate increases from 14.40\% to
19.62\%, with significant increases in 32 of 72 comparisons. The defence
therefore represents a robustness--false-positive trade-off rather than
complete prediction invariance. It substantially reduces harmful
vulnerable-to-benign transitions for classifiers such as the SVM, MLP,
and XGBoost, but provides weaker evidence for $k$-NN and can introduce
new missed-vulnerability and false-positive errors for individual
samples.

A further limitation is that the projector is trained jointly on all
four mutation families evaluated at test time. The reported results
therefore demonstrate robustness on unseen samples from known
transformation families, but do not establish generalization to unseen
mutation types. Future work should evaluate a zero-shot mutation setting
in which the projector is trained on only a subset of transformations,
such as comment removal and dead-code insertion, and then tested on
entirely held-out structural transformations. Additional work should
calibrate decision thresholds using validation data, evaluate other code
encoders and vulnerability datasets, and apply stronger behavioural
equivalence checks to the generated mutations.

\section{Acknowledgment} 
The authors express their sincere gratitude to Dr. Olga Vechtomova at the University of Waterloo for her inspiring instruction and guidance in Natural Language Processing. Her insightful feedback, encouragement, and dedication to research significantly shaped the ideas and development of this work.

\bibliographystyle{IEEEtran}
\bibliography{references}

@inproceedings{feng2020codebert,
  author    = {Zhangyin Feng and
               Daya Guo and
               Duyu Tang and
               Nan Duan and
               Xiaocheng Feng and
               Ming Gong and
               Linjun Shou and
               Bing Qin and
               Ting Liu and
               Daxin Jiang and
               Ming Zhou},
  title     = {{CodeBERT}: A Pre-Trained Model for Programming and Natural Languages},
  booktitle = {Findings of the Association for Computational Linguistics: EMNLP 2020},
  pages     = {1536--1547},
  publisher = {Association for Computational Linguistics},
  year      = {2020},
  doi       = {10.18653/v1/2020.findings-emnlp.139},
  url       = {https://doi.org/10.18653/v1/2020.findings-emnlp.139}
}

@inproceedings{karmakar2021pretrained,
  author    = {Anjan Karmakar and Romain Robbes},
  title     = {What Do Pre-Trained Code Models Know About Code?},
  booktitle = {Proceedings of the 36th IEEE/ACM International Conference on Automated Software Engineering},
  pages     = {1332--1336},
  publisher = {IEEE},
  year      = {2021},
  doi       = {10.1109/ASE51524.2021.9678927},
  url       = {https://doi.org/10.1109/ASE51524.2021.9678927}
}

@inproceedings{zhou2019devign,
  author    = {Yaqin Zhou and
               Shangqing Liu and
               Jingkai Siow and
               Xiaoning Du and
               Yang Liu},
  title     = {{Devign}: Effective Vulnerability Identification by Learning Comprehensive Program Semantics via Graph Neural Networks},
  booktitle = {Advances in Neural Information Processing Systems},
  volume    = {32},
  pages     = {10197--10207},
  publisher = {Curran Associates, Inc.},
  year      = {2019},
  url       = {https://proceedings.neurips.cc/paper/2019/hash/49265d2447bc3bbfe9e76306ce40a31f-Abstract.html}
}

@inproceedings{fan2020bigvul,
  author    = {Jiahao Fan and
               Yi Li and
               Shaohua Wang and
               Tien N. Nguyen},
  title     = {A {C/C++} Code Vulnerability Dataset with Code Changes and {CVE} Summaries},
  booktitle = {Proceedings of the 17th International Conference on Mining Software Repositories},
  pages     = {508--512},
  publisher = {Association for Computing Machinery},
  year      = {2020},
  doi       = {10.1145/3379597.3387501},
  url       = {https://doi.org/10.1145/3379597.3387501}
}

@inproceedings{fu2022linevul,
  author    = {Michael Fu and Chakkrit Tantithamthavorn},
  title     = {{LineVul}: A Transformer-Based Line-Level Vulnerability Prediction},
  booktitle = {Proceedings of the 19th International Conference on Mining Software Repositories},
  pages     = {608--620},
  publisher = {Association for Computing Machinery},
  year      = {2022},
  doi       = {10.1145/3524842.3528452},
  url       = {https://doi.org/10.1145/3524842.3528452}
}

@inproceedings{henkel2022semantic,
  author    = {Jordan Henkel and
               Goutham Ramakrishnan and
               Zi Wang and
               Aws Albarghouthi and
               Somesh Jha and
               Thomas W. Reps},
  title     = {Semantic Robustness of Models of Source Code},
  booktitle = {Proceedings of the IEEE International Conference on Software Analysis, Evolution and Reengineering},
  pages     = {526--537},
  publisher = {IEEE},
  year      = {2022},
  doi       = {10.1109/SANER53432.2022.00070},
  url       = {https://doi.org/10.1109/SANER53432.2022.00070}
}

@inproceedings{yang2022alert,
  author    = {Zhou Yang and
               Jieke Shi and
               Junda He and
               David Lo},
  title     = {Natural Attack for Pre-Trained Models of Code},
  booktitle = {Proceedings of the 44th International Conference on Software Engineering},
  pages     = {1482--1493},
  publisher = {Association for Computing Machinery},
  year      = {2022},
  doi       = {10.1145/3510003.3510146},
  url       = {https://doi.org/10.1145/3510003.3510146}
}

@inproceedings{na2023dip,
  author    = {CheolWon Na and
               YunSeok Choi and
               Jee-Hyong Lee},
  title     = {{DIP}: Dead Code Insertion Based Black-Box Attack for Programming Language Model},
  booktitle = {Proceedings of the 61st Annual Meeting of the Association for Computational Linguistics},
  pages     = {7777--7791},
  publisher = {Association for Computational Linguistics},
  year      = {2023},
  doi       = {10.18653/v1/2023.acl-long.430},
  url       = {https://doi.org/10.18653/v1/2023.acl-long.430}
}

@inproceedings{jain2021contrastive,
  author    = {Paras Jain and
               Ajay Jain and
               Tianjun Zhang and
               Pieter Abbeel and
               Joseph Gonzalez and
               Ion Stoica},
  title     = {Contrastive Code Representation Learning},
  booktitle = {Proceedings of the 2021 Conference on Empirical Methods in Natural Language Processing},
  pages     = {5954--5971},
  publisher = {Association for Computational Linguistics},
  year      = {2021},
  doi       = {10.18653/v1/2021.emnlp-main.482},
  url       = {https://doi.org/10.18653/v1/2021.emnlp-main.482}
}

@inproceedings{wang2022bridging,
  author    = {Deze Wang and
               Zhouyang Jia and
               Shanshan Li and
               Yue Yu and
               Yun Xiong and
               Wei Dong and
               Xiangke Liao},
  title     = {Bridging Pre-Trained Models and Downstream Tasks for Source Code Understanding},
  booktitle = {Proceedings of the 44th International Conference on Software Engineering},
  pages     = {287--298},
  publisher = {Association for Computing Machinery},
  year      = {2022},
  doi       = {10.1145/3510003.3510062},
  url       = {https://doi.org/10.1145/3510003.3510062}
}

@inproceedings{croft2023dataquality,
  author    = {Roland Croft and
               M. Ali Babar and
               M. Mehdi Kholoosi},
  title     = {Data Quality for Software Vulnerability Datasets},
  booktitle = {Proceedings of the 45th IEEE/ACM International Conference on Software Engineering},
  pages     = {121--133},
  publisher = {IEEE},
  year      = {2023},
  doi       = {10.1109/ICSE48619.2023.00022},
  url       = {https://doi.org/10.1109/ICSE48619.2023.00022}
}

@inproceedings{rahman2024causalvul,
  author    = {{Md Mahbubur Rahman} and
               Ira Ceka and
               Chengzhi Mao and
               Saikat Chakraborty and
               Baishakhi Ray and
               Wei Le},
  title     = {Towards Causal Deep Learning for Vulnerability Detection},
  booktitle = {Proceedings of the 46th IEEE/ACM International Conference on Software Engineering},
  pages     = {1888--1898},
  publisher = {Association for Computing Machinery},
  year      = {2024},
  doi       = {10.1145/3597503.3639170},
  url       = {https://doi.org/10.1145/3597503.3639170}
}

@inproceedings{ding2025primevul,
  author    = {Yangruibo Ding and
               Yanjun Fu and
               Omniyyah Ibrahim and
               Chawin Sitawarin and
               Xinyun Chen and
               Basel Alomair and
               David Wagner and
               Baishakhi Ray and
               Yizheng Chen},
  title     = {Vulnerability Detection with Code Language Models: How Far Are We?},
  booktitle = {Proceedings of the 47th IEEE/ACM International Conference on Software Engineering},
  pages     = {1729--1741},
  publisher = {IEEE},
  year      = {2025},
  doi       = {10.1109/ICSE55347.2025.00038},
  url       = {https://doi.org/10.1109/ICSE55347.2025.00038}
}

@article{jolliffe2016pca,
  author  = {Ian T. Jolliffe and Jorge Cadima},
  title   = {Principal Component Analysis: A Review and Recent Developments},
  journal = {Philosophical Transactions of the Royal Society A: Mathematical, Physical and Engineering Sciences},
  volume  = {374},
  number  = {2065},
  pages   = {20150202},
  year    = {2016},
  doi     = {10.1098/rsta.2015.0202},
  url     = {https://doi.org/10.1098/rsta.2015.0202}
}

@article{mcinnes2018umap,
  author  = {Leland McInnes and John Healy and Nathaniel Saul and Lukas Grossberger},
  title   = {{UMAP}: Uniform Manifold Approximation and Projection},
  journal = {Journal of Open Source Software},
  volume  = {3},
  number  = {29},
  pages   = {861},
  year    = {2018},
  doi     = {10.21105/joss.00861},
  url     = {https://doi.org/10.21105/joss.00861}
}

\end{document}